\documentclass[12pt]{article}   
\pdfoutput=1
\usepackage{geometry,bbm}   
\usepackage{ulem}
\usepackage[toc,page]{appendix}             		
\usepackage{graphicx}
\usepackage[usenames,dvipsnames]{color}
\usepackage{slashed}					
\usepackage{amssymb,amsmath}
\usepackage{hyperref}
\usepackage{mathtools}
\usepackage{cite}

\usepackage{graphicx}
\usepackage{bm}
\def\bea{\begin{eqnarray}}
\def\eea{\end{eqnarray}}
\usepackage{latexsym}
\usepackage{epsfig,amssymb,euscript}
\usepackage{amsmath}
\usepackage{array,calc,epsfig}

\usepackage{color}

\newcommand{\Tr}{{\rm Tr}}
\newcommand{\A}{{\cal A}}
\newcommand{\w}{\wedge}

\def\){\right)}
\def\({\left( }
\def\]{\right] }
\def\[{\left[ }

\newcommand{\be}{\begin{equation}}
\newcommand{\ee}{\end{equation}}

\def\bea{\begin{eqnarray}}
\def\eea{\end{eqnarray}}

\def\bal#1\eal{\begin{align}#1\end{align}}

\def\bald{\begin{aligned}}
\def\eald{\end{aligned}}

\def\beqx{\begin{displaymath}}
\def\eeqx{\end{displaymath}}

\newcommand{\bmat}{\left(\begin{array}}
\newcommand{\emat}{\end{array}\right)}

\def\bo{{\raise-.3ex\hbox{\large$\Box$}}}               
\def\face{{\raise.2ex\hbox{$\displaystyle \bigodot$}\mskip-2.2mu \llap {$\ddot
        \smile$}}}                                   
\def\>{\rangle}                                      
\def\<{\langle}                                      

\def\leftrightarrowfill{$\mathsurround=0pt \mathord\leftarrow \mkern-6mu
        \cleaders\hbox{$\mkern-2mu \mathord- \mkern-2mu$}\hfill
        \mkern-6mu \mathord\rightarrow$}        
\def\dvec#1{\vbox{\ialign{##\crcr
        \leftrightarrowfill\crcr\noalign{\kern-1pt\nointerlineskip}
        $\hfil\displaystyle{#1}\hfil$\crcr}}}           
\def\Tr{{\rm Tr \,}}                                    

\def\-{\hphantom{-}}

\numberwithin{equation}{section}

\begin{document}

\begin{titlepage}

\begin{flushright}
SISSA 24/2018/FISI
\end{flushright}
\bigskip
\def\thefootnote{\fnsymbol{footnote}}

\begin{center}
{\LARGE
{\bf
QCD domain walls, Chern-Simons \vskip 14pt theories and holography
}}
\end{center}

\bigskip
\begin{center}
{\large
Riccardo Argurio$^{1}$, Matteo Bertolini$^{2,3}$, Francesco Bigazzi$^{4}$, \\ \vskip 5pt Aldo L. Cotrone$^{4,5}$, Pierluigi Niro$^{1,6}$}

\end{center}

\renewcommand{\thefootnote}{\arabic{footnote}}

\begin{center}
\vspace{0.2cm}
$^1$ {Physique Th\'eorique et Math\'ematique and International Solvay Institutes \\ Universit\'e Libre de Bruxelles, C.P. 231, 1050 Brussels, Belgium\\}
$^2$ {SISSA and INFN - 
Via Bonomea 265; I 34136 Trieste, Italy\\}
$^3$ {ICTP - 
Strada Costiera 11; I 34014 Trieste, Italy\\}
$^4$ {INFN, Sezione di Firenze; Via G. Sansone 1; I-50019 Sesto Fiorentino (Firenze), Italy\\}
$^5$ {Dipartimento di Fisica e Astronomia, Universit\'a di Firenze; Via G. Sansone 1; I-50019 Sesto Fiorentino (Firenze), Italy\\}
$^6$ {Theoretische Natuurkunde, Vrije Universiteit Brussel\\ Pleinlaan 2, 1050 Brussels,
Belgium\\}
\vskip 5pt
{\texttt{rargurio@ulb.ac.be, bertmat@sissa.it, bigazzi@fi.infn.it, cotrone@fi.infn.it, pierluigi.niro@ulb.ac.be}}

\end{center}

\vskip 5pt
\noindent
\begin{center} {\bf Abstract} \end{center}
\noindent
Massive QCD at $\theta=\pi$ breaks CP spontaneously and admits domain walls whose dynamics and phases depend on the number of flavors and their masses. We discuss these issues within the Witten-Sakai-Sugimoto model of holographic QCD. Besides showing that this model reproduces all QCD expectations, we address two interesting claims in the literature. The first is about the possibility that the QCD domain-wall theory is fully captured by three-dimensional physics, only. The second regards the existence of quantum phases in certain Chern-Simons theories coupled to fundamental matter. Both claims are supported by the string theory construction.

\vspace{1.6 cm}
\vfill

\end{titlepage}

\newpage
\tableofcontents

\section{Introduction}
\label{intro}

One interesting conjecture about the dynamics of massive QCD is the existence of dynamical domain walls (DWs) at $\theta=\pi$ separating two gapped vacua where CP symmetry is spontaneously broken. The existence and the dynamics of these DWs depend on the number of flavors $N_f$ and their masses.  Recently, a rather coherent picture of this phenomenon has been provided \cite{Gaiotto:2017tne}, based on previous large $N$ results, anomaly inflow arguments and known facts about ${\cal N}=1$ Super-Yang-Mills.  

Let us summarize the main aspects of this picture. 

When all quarks are massive, the QCD $\theta$ angle becomes physical, like in pure Yang-Mills (YM). Let us pick for simplicity the degenerate case where all the flavors have the same mass $m$. Because of the chiral anomaly, the physics does not depend separately on $m$ and $\theta$ but rather on the complex combination $m^{N_f} e^{i \theta}$. Hence, following \cite{Gaiotto:2017tne}, we will consider $m$ real and positive in the following. Thereafter we refer, for definiteness, to an $SU(N)$ gauge group.

For $N_f=1$ spontaneous breaking of CP (and hence the DW) at $\theta=\pi$ exists only from some positive value of the mass $m_0$ on. For very large mass the theory reduces to pure YM and the DW theory is, like for pure YM, a Chern-Simons (CS) $SU(N)_1$ theory plus the trivial center of mass mode. This phase does not persist if one lowers the mass. For $m = m_{tr} \sim \Lambda$ (with $\Lambda$ the QCD dynamical scale) the DW undergoes a phase transition towards $SU(N)_0$ CS theory, which is a trivial theory. Finally, for even smaller mass, smaller than $m_0\sim \Lambda/N$, the DW ceases to exist. At $m=m_0$  the phase transition in the bulk  becomes second order, with the only massless particle being the $\eta^\prime$ ({\it i.e.}~the would-be Goldstone boson associated to the axial $U(1)$ symmetry). 

For $N_f>1$ the story is quite different. First, the DW exists for any non-vanishing value of quark masses. The theory at $m=0$ is nothing but the $\sigma$-model chiral symmetry breaking phase of massless QCD, described by the chiral Lagrangian. For large enough mass the DW dynamics reduces, again, to that of the pure YM domain wall, namely an $SU(N)$ gauge theory at CS level $k=1$. Lowering the mass, the DW theory undergoes a phase transition and, for $m< m_{tr}$, it enters a new phase described by a $\mbox{CP}^{N_f-1}$ non-linear $\sigma$-model (which, for small enough mass, can be seen directly in the chiral Lagrangian).  

Finally, for $N_f > N_{CFT}$, with $N_{CFT}$ being some function of $N$, massless QCD is conjectured to flow in the IR to an interacting fixed point and, for even larger $N_f$, it becomes IR free. In this regime, when quark masses are turned on, the domain-wall theory is argued not to undergo a phase transition and it is believed to be described by $SU(N)_1$ CS theory for any value of the mass. We will not consider this regime further in this paper, as we will always be in the limit in which $N_f$ is parametrically smaller than $N$. 

It must be stressed that when the phase transition occurs on the DW, the four-dimensional physics,  both for $N_f=1$ and $N_f > 1$, is instead completely smooth as one varies $m$ through $m_{tr}$. An interesting aspect, which will be relevant later, regards the DW dynamics {\it per se}. If the phase transition at $m = m_{tr}$ is second order, being the bulk gapped, the DW dynamics around the transition should be solely described in terms of a three-dimensional theory. Relying on the analysis in \cite{Komargodski:2017keh}, this was proposed in \cite{Gaiotto:2017tne} to be an $SU(N)_{1 - N_f/2}$ CS theory coupled to $N_f$ fundamental fermions which, via boson/fermion duality \cite{Aharony:2015mjs}, is dual to $U(1)_{-N}$ CS theory coupled to $N_f$ fundamental (Wilson-Fisher) scalars. Similarly, in the neighborhood of the phase transition, where three-dimensional matter is massive, level-rank duality would be at work.

\begin{figure}[ht]
\begin{center}
\includegraphics[height=0.20\textheight]{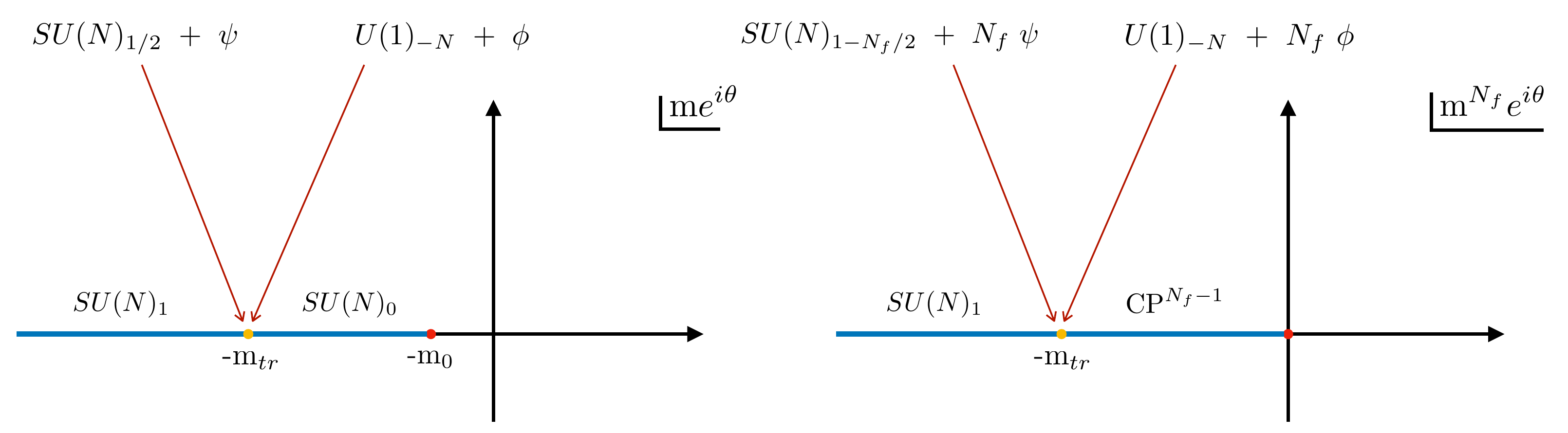}
\caption{\small{Domain walls of massive QCD at $\theta=\pi$ and their phases in the complex $m^{N_f} e^{i\theta}$ plane, including the proposed three-dimensional CS theories describing them. On the left, QCD with $N_f=1$; on the right, QCD with $1 < N_f < N_{CFT}$. The blue line represents the values of quark masses for which the domain wall exists. At $m_{tr} \sim \Lambda$ a phase transition occurs. The DW ceases to exist at $m=m_0$ and $m=0$ for $N_f=1$ and $N_f>1$, respectively.}}
\label{GKS}
\end{center}
\end{figure}

Figure \ref{GKS} summarizes the DW theory for massive QCD in the two qualitatively different regimes of interest. We refer to \cite{Gaiotto:2017tne} and references therein for more details (see also the other recent papers \cite{Gaiotto:2017yup,DiVecchia:2017xpu}). 

We would like to analyze the phases of massive QCD in the context of the Witten-Sakai-Sugimoto model \cite{Witten:1998zw,Sakai:2004cn} (WSS for short); to date, the most successful top-down holographic model of large $N$  QCD. 

There are at least two motivations behind our analysis. On the one hand, that the WSS model may reproduce a very non-trivial aspect of massive QCD dynamics, such as its domain walls, would give further evidence that it is in the same universality class of QCD. On the other hand, it is often the case that having a string embedding of QFT phenomena sheds light on aspects which are not fully under control within the QFT framework. For example, varying quark masses, the DW theory should undergo a phase transition as we reviewed above, but present field theory tools do not allow to conclude about the order of such phase transition, nor whether the transition can be fully captured by three-dimensional physics, only.\footnote{For $SU(N)_k$ CS theory with fermionic matter and its bosonic dual, in the limit $N,k\gg 1$ with $N/k$ fixed, arguments such as those in \cite{Giombi:2011kc,Aharony:2011jz,Choudhury:2018iwf} indicate the existence of a CFT at the critical point.} As we will show, WSS open string dynamics suggests that the phase transition can indeed be fully captured in terms of a three-dimensional CS theory, giving support to the claims in \cite{Gaiotto:2017tne} as well as those in  \cite{Komargodski:2017keh}.

Let us anticipate, in a nutshell, the picture that will emerge. In the WSS model $N$ wrapped D4-branes support the Yang-Mills sector, while $N_f$ D8-branes account for the quark content. The DW at $\theta=\pi$ is associated to a D6-brane, which naturally has a $U(1)_N$ CS theory on its world-volume, with $N_f$ charged scalars if there are $N_f$ flavors. These scalars are the would-be tachyons between the DW D6-brane  and the flavor D8-branes. Above the transition at $m_{tr}$, the scalars have a positive squared mass and the D6-brane is a physical object in the set-up. Below the transition, when tachyons condense, instead, the D6-brane dissolves into flux inside the D8-branes. In this regime the low-energy picture from WSS reduces exactly to the one of the chiral Lagrangian (including also the $\eta'$, as we are in the large $N$ limit), with the same symmetry breaking pattern and $\sigma$-model description.

The rest of the paper is organized as follows. In section \ref{revWSS} we review the basics of the WSS construction, emphasizing those aspects which are relevant for the present  discussion.  In section \ref{YMdw} we discuss, within the WSS model, domain walls in the pure Yang-Mills case and move on in section \ref{mQCD} to the case of QCD. A final section contains a summary of our results and an outlook.

\section{The Witten-Sakai-Sugimoto model}
\label{revWSS}

The Witten-Sakai-Sugimoto (WSS) model \cite{Witten:1998zw,Sakai:2004cn} is a non-supersymmetric gauge theory realized on intersecting D-branes in type IIA string theory, whose deep IR dynamics reduces to that of (large $N$) QCD. The set-up consists of $N$ D4-branes wrapping a circle $S_{x_4}$ of radius $M_{KK}^{-1}$ and two stacks of $N_f$ D8 and $\overline{\mbox{D8}}$-branes localized at antipodal points on that circle.\footnote{Non-antipodal configurations generate spurious quartic fermionic interaction terms  \cite{kutnjl}.} With periodic (resp. antiperiodic) boundary conditions for bosons (resp. fermions) on the circle, the dynamics on the D4-branes at energies $E\ll M_{KK}$ is described by a pure $U(N)$ gauge theory in $3+1$ dimensions, as all the (Kaluza-Klein) adjoint matter fields get masses of the order of $M_{KK}$. The low-energy modes from D4/D8 and D4/$\overline{\mbox{D8}}$ open strings are $N_f$ chiral fundamental fermions, {\it i.e.} massless quarks.  

In the 't Hooft large $N$ limit where quarks can be treated in the quenched approximation\footnote{See \cite{sonnesme,noiWSSsmeared} for results including the flavor backreaction to leading order in $N_f/N$.} the holographic dual of the interacting $SU(N)$ sector of the model is a regular gravity background \cite{Witten:1998zw} (henceforth, WYM background, see appendix \ref{appreview} for more details), with ${\mathbb R}^{1,3} \times\rm{Disk} \times S^4$ topology, probed by the flavor D8-branes.   

Let us first focus on the pure glue sector. The background includes a running dilaton and $N$ units of RR four-form $F_4=dC_3$ flux through $S^4$
\be
\int_{S^4} F_4 = 8\pi^3 \alpha'^{3/2} g_s N\,.
\label{flux}
\ee 
The disk topology is that of a cigar surface designed by the $S_{x_4}$ circle smoothly shrinking to zero size as the transverse radial coordinate (holographically mapped into the field theory RG scale) reaches its minimal value. 

Holographic computations show that the Yang-Mills theory dual to such a background has a mass gap and confines. The glueball mass scale and therefore $\Lambda$, is given by $M_{KK}$. The string tension scales like $\lambda M_{KK}^2$ where $\lambda\equiv g_{YM}^2N = 2\pi g_s N \alpha'^{1/2} M_{KK}$. Reliability of the gravity description requires $\lambda\gg1$, which means that the spurious Kaluza-Klein modes are actually not decoupled from the Yang-Mills sector in this limit, as usual in holographic duals of confining theories in the supergravity regime. 

The YM $\theta$ angle, which will play a key role in our analysis, can be turned on in the model by including a RR one-form potential $C_1$ such that
\be   
\theta + 2\pi k=\frac{1}{g_s \sqrt{\alpha'}}\int_{S_{x_4}} C_1~~, ~~ k \in {\mathbb Z}\,,
\ee
where the integral is evaluated at the boundary \cite{Witten:1998uka}. When $\lambda\theta/N\ll1$ the backreaction of $C_1$ on the above background can be neglected. This is the limit in which we will work.\footnote{See, {\it e.g.}, \cite{dubo,wymtheta} and references therein for an analysis of the case with backreacting $C_1$. Most importantly, in this case one can verify that WYM has a mass gap also at finite $\theta$.} 

Domain walls at $\theta=\pi$ are described by D6-branes sitting at the tip of the cigar, extending on $2+1$ directions in $ {\mathbb R}^{1,3}$ (say $x_0, x_2, x_3$) and wrapping the four-sphere $S^4$. D6-branes couple (magnetically) to $C_1$ and crossing a D6-brane the $\theta$ angle gets shifted by one, in units of $2\pi$. So, at $\theta=\pi$, D6-branes interpolate between $\theta=-\pi$ and $\theta=\pi$ vacua.

\begin{figure}[ht]
\begin{center}
\includegraphics[height=0.20\textheight]{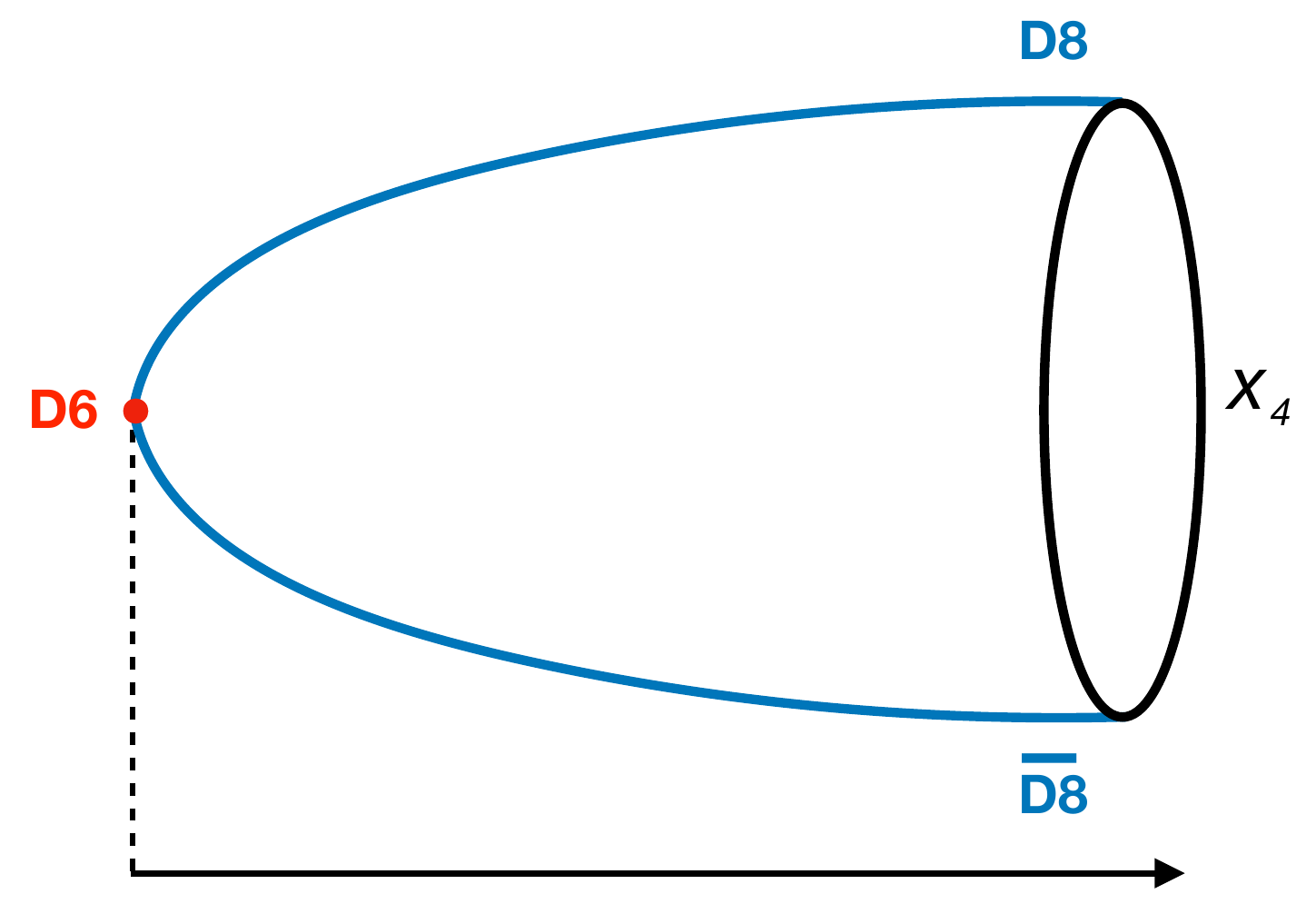}
\caption{{\small Flavors and domain walls in the Witten-Sakai-Sugimoto model. The flavor D8/$\overline{\mbox{D8}}$-branes are placed at fixed antipodal point on the $S_{x_4}$ circle and extend in the transverse radial direction. The D6-branes sit at the tip of the cigar geometry.}}
\label{WSS}
\end{center}
\end{figure}

The $U(N_f)_L\times U(N_f)_R$ gauge symmetry on the $N_f$ D8/$\overline{\mbox{D8}}$-branes corresponds to the classical global symmetry of the dual QCD-like theory. On the background above, the flavor branes minimize their energy taking a U-shaped configuration along the cigar, gluing up at the tip. This geometrically realizes the spontaneous chiral symmetry breaking $U(N_f)_L\times U(N_f)_R\rightarrow U(N_f)$ in the dual gauge theory \cite{Sakai:2004cn}. Figure \ref{WSS} displays the cigar geometry and the embedding of the D8 flavor branes and D6 domain walls therein.  

The hadronic sector of the WSS model with massless quarks is described by the effective action for the U-shaped D8-branes. At leading order in $\alpha'$ and after reduction on $S^4$ the latter is the action for a $U(N_f)$ gauge field ${\cal A}$, with Chern-Simons terms, on a curved $4+1$ dimensional background
\begin{equation}\label{SYM}
S_{WSS} = -\kappa\int d^4x d z\, \left(\frac{1}{2}k(z)^{-1/3}  \, \Tr\mathcal{F}_{\mu\nu}\mathcal{F}^{\mu\nu} +   k(z)  \Tr\mathcal{F}_{\mu z}\mathcal{F}^\mu_{\;\; z}\right)
 + \frac{N}{24\pi^2}\int  \omega_5(\mathcal{A})\,,
\end{equation}
where $\kappa \sim NM_{KK}^2$ ($\lambda$ dependence is understood from now on, except when it will play a relevant role), $k(z) = (1+z^2)$,
$d \omega_5(\mathcal{A}) = \Tr\mathcal{F}^{  3}$ and $z\in (-\infty,\infty)$ is the dimensionless holographic radial coordinate ($z=0$ corresponding to the tip of the cigar).

Mesonic states in the gauge theory correspond to fluctuations of the gauge field ${\cal A}$, while instantonic configurations are interpreted as baryons. The pseudoscalar mesons (which, at large $N$, have all the same decay rate $f_{\pi}$) are described by the holonomy matrix
\begin{equation}
{\cal U}(x^\mu) = \mathcal{P} \exp\left( -i \int_{-\infty}^\infty d z\, \mathcal{A}_z(x^\mu,z)\right)=\exp \left(\frac{2i}{f_\pi}\pi^a(x^\mu) T^a\right)\,,
\label{pionm}
\end{equation}
where $T^a$ are $U(N_f)$ generators normalized as $\Tr(T^aT^b)=\frac{1}{2}\delta_{ab}$ and $f^2_{\pi}=4\kappa/\pi $.
Clearly, as in QCD, the low-energy physics is dominated by pseudoscalar Goldstone bosons, since all the other modes (glueballs, KK modes and vector mesons) have masses of order $M_{KK}$. The D8 effective action above, restricted to the pseudoscalar meson sector, precisely reduces to the chiral Lagrangian. 

Turning on both flavor masses and a finite $\theta$ angle \cite{nedmshort,Bartolini:2016dbk}, amounts to adding to the action (\ref{SYM}) the terms 
\be\label{Smass}
S_{\mathrm{mass}} = c \int d^4x\, \Tr\mathcal{P}\left[M e^{-i\int \mathcal{A}_z dz}+ \mathrm{h.c.}\right]\,, 
\ee
where (according to the results in \cite{McNees:2008km}) $c\sim N M_{KK}^3$ and
\be\label{SF2}
S_{\theta} = -\frac{\chi_g}{2}\int d^4x\left(\theta + \int dz \Tr\mathcal{A}_z\right)^2\,,
\ee
where $\chi_g\sim M_{KK}^4$ is the topological susceptibility of the unflavored theory.

The mass term (\ref{Smass}), where $M$ is the quark mass matrix, arises from the one-instanton contribution of long open strings attached to the antipodal flavor branes \cite{AK,Hashimotomass}.\footnote{See \cite{Bergman:2007pm,Dhar:2007bz,Dhar:2008um} for an alternative proposal to introduce masses in the WSS model.} 
Due to the relation (\ref{pionm}), it is easy to realize that eq. (\ref{Smass}) has precisely the same structure of the mass term $\Tr [M\mathcal{U}+\mathrm{h.c.}]$ introduced in the chiral Lagrangian approach. In the following we will focus on the mass-degenerate case $M_{ij} = m \,\delta_{ij}$, choosing $m$ to be real. The one-instanton approximation is reliable if $m\ll M_{KK}$ \cite{AK, Hashimotomass}.

The term $S_{\theta}$ (\ref{SF2}) is induced by the anomaly inflow on $dC_1$ which ceases to be gauge invariant in presence of the D8-branes \cite{Sakai:2004cn, Bartolini:2016dbk}. In absence of flavor mass terms, the $\theta$ term can be removed  via a gauge shift of the abelian component $\widehat A_z$ of the $U(N_f)$ gauge field \cite{Bartolini:2016dbk}.\footnote{Here we decompose the $U(N_f)$ gauge field as $\mathcal{A} =  \mathbf{1} (2N_f)^{-1/2}\widehat{A} + A^a t^a$, where $t^a$ are the $SU(N_f)$ generators. } This is precisely the way in which holography accounts for the fact that in the presence of (even just one) massless flavors, the $\theta$ dependence is washed away by a chiral rotation of the quark fields. If all quarks are massive, instead, the $\theta$ parameter cannot be gauged away, and becomes physical. In this case the domain walls at $\theta=\pi$ are again expected to be related to D6-branes sitting at the tip of the cigar.

\section{Domain walls of pure Yang-Mills}
\label{YMdw}

In this section we want to investigate the DW theory of planar $SU(N)$ Yang-Mills theory by analyzing the IR of its gravitational dual provided by Witten \cite{Witten:1998zw}.\footnote{See \cite{Acharya:2001dz} as the first instance of domain-wall theories in a holographic context, where the $k$-BPS domain walls of ${\cal N} = 1$ pure SYM have been analyzed.} Besides being interesting {\it per se}, this will also be an important ingredient when considering the addition of flavors. In particular, for large enough quark masses QCD should behave as pure YM at low energy, and the corresponding DW dynamics should then reduce to that of pure YM DWs.

As already mentioned, in WYM domain walls are described by D6-branes wrapped on the four-sphere \cite{Witten:1998uka}. The theory on the DW is readily seen to be $U(1)_{N}$ (plus the center of mass mode, given by the massless world-volume scalar corresponding to the transverse $x_1$ flat direction). Indeed, the DBI action of a D6-brane gives, upon integration on the four-sphere, the three-dimensional gauge field kinetic term while the WZ term reads
\be
\label{wzk1}
\tau_6 \frac{(2\pi \alpha')^2}{2} \int_{M_7} C_3 \wedge {\cal F} \wedge {\cal F} =  \tau_6 \frac{(2\pi \alpha')^2}{2} \int_{S^4 \times M_3} F_4 \wedge {\cal A}\wedge {\cal F} = \frac{N}{4\pi} \int_{M_3} {\cal A} \wedge {\cal F} ~,
\ee
where ${\cal F}$ is the gauge field strength on the D6 and, in the last equality, we have used eq.~(\ref{flux}).\footnote{We use conventions where 
$\tau_p=(2\pi)^{-p} \alpha'^{-(p+1)/2}g_s^{-1}$.} By level-rank duality, this is dual to $SU(N)_{-1}$, thus reproducing the expected behavior of pure YM DWs (with respect to the conventions of \cite{Gaiotto:2017tne} used in the introduction, we now consider for convenience  the time-reversed situation). 

In this respect, a comment is in order. In \cite{Gaiotto:2017yup} it has been argued that the DW of pure YM supports a non-spin $SU(N)_{-1}$ theory, which is not level-rank dual to $U(1)_{N}$ (level-rank duality holds between spin theories only; in particular, $U(1)_{N}$ for $N$ odd does not exist as a non-spin theory). In the model at hand YM has a UV completion involving fermionic degrees of freedom and the theory automatically includes a choice of spin structure. The DW theory in WYM is thus a spin CS theory, and level-rank duality between $U(1)_N$ and $SU(N)_{-1}$ is well defined. This implies that, strictly speaking, WYM and pure YM theories are not in the same universality class, at least in the sense of \cite{Gaiotto:2014kfa}, since they have a different spectrum of line operators on the DWs. Note, however, that in the same sense pure YM is not in the same universality class of QCD with massive quarks, or softly broken ${\cal N}=1$ SYM, which are instead in the same universality class of the WSS model.

Let us take a closer look at the WZ term \eqref{wzk1} and consider, more generally,  the effective action of $k$ coincident D6-branes (corresponding to a jump of $2\pi k$ of the $\theta$ angle). Their WZ action, including the relevant part of the A-roof genus, reads (see, {\it e.g.}, \cite{Johnson:2003gi} and references therein)
\be
\tau_6 \int_{M_7} C_3 \wedge {\rm Tr}_k \,e^{2\pi \alpha'{\cal F}+B}\left(1-\frac{p_1(4\pi^2 \alpha' R)}{48} \right)~, \label{wzaroff}
\ee
where $p_1(R)$ is the first Pontryagin class
\be
p_1(R)=- \frac{1}{8\pi^2} {\rm Tr} \,R\wedge R~.
\ee
Thus, in the WYM background there can be all of the following terms 
\bea
&& \tau_6 \frac{(2\pi \alpha')^2}{2} \int_{M_7} C_3 \wedge {\rm Tr}_k \,{\cal F} \wedge {\cal F} = \frac{N}{4\pi} \int_{M_3} {\rm Tr}_k \left( {\cal A} \wedge d{\cal A} -  \frac23 i \, \A \w \A \w \A \right) \label{cs}\\
&& \tau_6 (2\pi \alpha') \int_{M_7} C_3 \wedge {\rm Tr}_k \,{\cal F} \wedge B = \frac{N}{4\pi} \int_{M_3} \frac{2}{2\pi\alpha'} {\rm Tr}_k\,\A \wedge B \label{extra}\\
&& \tau_6 \frac{(4\pi^2 \alpha')^2}{8\cdot 48 \pi^2} \int_{M_7} C_3 \w {\rm Tr} R\wedge R = N k\int_{M_3} CS_g~, \label{csg}
\eea
where
\be
\int_{M_3= \partial X_4} CS_g = \frac{1}{192 \pi} \int_{X_4}  {\rm Tr}\, R\wedge R 
\ee
is the gravitational CS term related to the framing anomaly.\footnote{Note that from the point of view of the DW theory, $R$ and $B$ are background gauge fields.} 
The action given by the sum of (\ref{cs})-(\ref{csg}) is precisely the one of a $U(k)_N$ CS theory in presence of an external two-form potential for the center one-form symmetry \cite{Hsin:2016blu,Benini:2017aed}.\footnote{We adhere to standard notation where
\bea
U(k)_{N,M} =  \frac{SU(k)_N \times U(1)_{kM}}{Z_k}~~~\mbox{and}~~~U(k)_N \equiv U(k)_{N,N}~. \nonumber
\eea
} In particular, the term (\ref{extra}) is the one needed to preserve gauge invariance on the domain wall, when we couple the 4d theory with the $B$ field two-form, see, {\it e.g.}, \cite{Dierigl:2014xta}. The mixed anomaly between the center $Z_N$ and CP symmetries of the 4d theory at $\theta=\pi$ should be matched by the 3d theory \cite{Gaiotto:2017yup}. 
This dictates that the 3d theory cannot be trivial and  it must acquire a counterterm upon a CP transformation. The counterterm is precisely  (\ref{extra}).

Note that the numerical factor of the term \eqref{csg} is half the one that usually appears in level-rank $U/SU$ dualities, see {\it e.g.} \cite{Hsin:2016blu,Benini:2017aed}.  
This, in principle, is not a problem, since $R$ is an external field: it just means that the dual $SU(N)$ theory must come with the same term \eqref{csg}, but with opposite sign.\footnote{We thank F. Benini for a discussion on this point.} 

In the present D-brane setting, this can be explicitly checked in the following way. 
As in an ordinary field theory set-up \cite{Gaiotto:2017yup}, a DW (let us consider $k=1$ now) can be artificially produced considering a non-dynamical deformation of the theory on $N$ wrapped D4-branes where  $\theta$ jumps from $-\pi$ to $\pi$ at some location, say at $x_1=0$ 
\be
\frac{1}{g_s \sqrt{\alpha'}}\int_{S_{x_4}} C_1 = \theta(x_1) = -\pi+2\pi\Theta(x_1)\,.
\ee
Considering the WZ term for the $N$ D4-branes, analogous to eq.~\eqref{wzaroff}, and integrating by parts, one immediately realizes that the above deformation is equivalent to adding to the 4d effective action on the branes (at constant $\theta=\pi$) an $SU(N)_{-1}$ Chern-Simons action on the wall. In principle, the low-energy degrees of freedom on the DW can be totally different from this UV CS theory. However, the 't Hooft anomalies of the IR theory on the DW have to match those of the UV CS action. It is thus very reasonable to assume that the effective action on the DW is actually given by an $SU(N)_{-1}$ CS theory \cite{Gaiotto:2017yup}, with the same anomaly terms as those deduced from the WZ action. 

Remarkably, in the present set-up this is much more than an assumption. Using holography, we have found that the DW theory is exactly the $U(1)_N$ level-rank dual theory. Coming back to our previous remark, one can now check that the anomalous gravitational coupling on the D6-brane corresponds, as expected, to an equal and opposite term in the $SU(N)_{-1}$ theory. We believe that this is what happens in any brane realization of level-rank duality, as can be checked for instance in the set-ups of \cite{Fujita:2009kw,Jensen:2017xbs}.

\section{Domain walls of massive QCD}
\label{mQCD}

In this section, by means of the WSS model, we would like to analyze the DW theories of planar, massive QCD at $\theta=\pi$ as a function of quark masses $m$ and number of flavors $N_f$ (as already mentioned, we take the quark masses to be degenerate, for simplicity). For large enough quark masses the theory should reduce, at low energy, to the one discussed in the previous section and so should the DW theory. Instead, for small masses,   eventually smaller than $\Lambda$, one expects to end up in the chiral Lagrangian regime. Following the discussion in the introduction, the DW theory should then undergo a phase transition and enter in a $\mbox{CP}^{N_f-1}$ $\sigma$-model phase or, for $N_f=1$, become a trivial theory. 

It is well-known that the WSS model, with small quark masses implemented as in \cite{AK, Hashimotomass}, does reproduce the chiral Lagrangian (including also the $\eta'$ particle) \cite{Sakai:2004cn} (see 
\cite{Bartolini:2016dbk} for the case of non-vanishing $\theta$, which is of our interest here). As such, following exactly the same steps of \cite{Gaiotto:2017tne}, it does also reproduce the low mass phase of both $N_f=1$ and $N_f>1$ QCD DWs (right parts of the two phase diagrams of figure \ref{GKS}). 

Below, we would like to emphasize some microscopic aspects of the WSS construction which will enable us to gain some intuition about DW dynamics outside the small mass regime; in particular, what the effective three-dimensional theory could be, if any. Although we can just be qualitative, a very natural and coherent picture emerges, which exactly agrees with the proposal of \cite{Gaiotto:2017tne}, including the proposed CS theory living on the domain walls.

As in the pure YM case, domain walls are described by D6-branes. The pure gauge sector on the D6-branes is the same as that discussed in the previous section. However, the presence of the D8-branes gives rise to a new sector of D6/D8 open strings, which in the low-energy limit are just $N_f$ scalar fields in the fundamental representation of the gauge group on the D6-branes. This shows that the field theory spectrum is exactly what predicted within the field theory analysis, see figure  \ref{GKS}.

At face value, these scalar fields are tachyonic. The D6-branes lie totally inside the D8-branes.  It is well-known that such a D$p$/D$(p+2)$ configuration has tachyons  in its $(p+1)$-dimensional world-volume. This signals the tendency of the branes to form a D$p$/D$(p+2)$ bound state, with energy lower than the sum of the energies of the respective constituent branes. 

In a general D$p$/D$(p+2)$ configuration, it is possible to keep the D$p$-branes as physical objects by holding them at a distance from the D$(p+2)$-branes large enough so that the tension of the strings compensates the tachyon negative zero-point energy (for this to happen, branes need to have boundary conditions keeping them apart, otherwise they will be attracted to each other). 

Closer to our set-up is a T-dual configuration, in which the D$p$ and D$(p+2)$ exchange roles, but they are no longer separated. What gives a positive mass shift to all D$p$/D$(p+2)$ modes is, now, a Wilson line (or, more precisely, a constant gauge field) on the D$(p+2)$ world-volume. In our set-up, it is precisely the presence of a non-zero $\theta$ angle that produces such a non-vanishing gauge field profile.

We now see what the effect of quark masses can be on the D6-branes. The D6-branes sit at the tip of the cigar, embedded in the D8-branes. At $\theta=\pi$, the latter have a non-vanishing gauge field along the $z$ direction, which has a non-zero and approximately constant value near the tip. It thus looks exactly like a flat-space constant gauge field, from the point of view of the D6-branes. As we will explicitly show in the next section, increasing quark masses the gauge field flux does increase, too. As a consequence, the D6/D8 modes get a positive mass shift, which increases, as quark masses are increased. So, in principle, for sufficiently strong gauge field profile, the lowest lying modes become a set of $N_f$ scalars with positive mass square, exactly as conjectured in \cite{Gaiotto:2017tne} for the large mass phase. 

Conversely, as quark masses are lowered, the corresponding flux on the D8-branes decreases, until when, below some critical value, the mass of the scalars becomes tachyonic. At this stage the tachyons condense, signalling that the branes have formed a bound state. It is given by the D8s with a world-volume flux $\widehat{F}$ turned on, such that it carries a D6 charge. 
In flat space, such a magnetic world-volume flux is completely delocalized \cite{Gava:1997jt,Aganagic:2000mh}.
In our set-up, the directions transverse to the D6 are $x_1$ and $z$, so the flux should be a $\widehat{F}_{x_1 z}$, such that
\be\label{D6inD8}
\int \widehat{F} \wedge C_7 \sim \int C_7~,
\ee
{\it i.e.}~a D6-brane charge for the bulk fields. The DWs at $\theta=\pi$ can now be seen purely within the world-volume theory of the D8-branes, which itself reproduces the physics of the chiral  Lagrangian, in agreement with \cite{Gaiotto:2017tne}, from which the DW theory is determined to be a $\sigma$-model (or a trivial theory with just the center of mass degree of freedom for $N_f=1$). This is interpreted as descending from the Higgsing of the $U(1)_{N}$ plus $N_f$ scalars (or, from boson-fermion duality, from $SU(N)_{-1}$ with $N_f$ fermions and a fermion condensate \cite{Komargodski:2017keh}).
Indeed, we can see the dissolution of the D6 into the D8s through tachyon condensation as a Higgsing process \cite{Gava:1997jt,Aganagic:2000mh}.

To summarize, the brane theory appears to realize precisely the scenario of the $U(1)_{N}$ CS theory plus $N_f$ scalars on the DW.

One can try to be more precise concerning the theory at the transition point. The $N_f$ tachyonic D6/D8 strings are a more prosaic realization of the Wilson-Fisher scalars of the conjectured DW theory. That their potential also includes a positive quartic term is evident from the fact that there is an endpoint to their condensation. Thus, at least in the crude limit in which we are considering them, they seem to reproduce a simple Landau-Ginzburg-like theory (with a CS term though), with the transition towards the symmetry-breaking phase taking place smoothly when their mass decreases to zero. In this respect, we seem to gather evidence for a phase transition of second order. This in turn would imply a transition point with rich conformal dynamics for gapless degrees of freedom confined to the DW. 

In the rest of this section we substantiate the picture outlined above with details from the WSS set-up. Note that while the D-brane construction clearly indicates that the DW is indeed a genuine three-dimensional theory, as for tachyon dynamics one can only evaluate the effects in the small quark mass regime, where the WSS model is reliable. In particular, we do not have direct access to the phase transition. This, eventually, does not let one give a clear-cut proof about the order of the phase transition. We will come back to this and some related aspects in section \ref{phtrs}. Below, we start focusing on the small mass regime, instead. We will mostly consider the $N_f=1$ case, since it already captures the bulk picture of the two phases, but we will also comment on $N_f>1$, when needed. The analysis of the vacua and the presence of DWs at $\theta=\pi$ has a long history, starting with \cite{Rosenzweig:1979ay,DiVecchia:1980yfw,Witten:1980sp,Nath:1979ik} and \cite{Kawarabayashi:1980dp, Ohta:1981ai}, to \cite{Evans:1996eq,Smilga:1998dh,Tytgat:1999yx} and the most recent \cite{Gaiotto:2017tne,DiVecchia:2017xpu,Cherman:2017dwt,Draper:2018mpj,Ritz:2018mce,Aitken:2018mbb}.

\subsection{Small quark mass regime}

We start by considering the WSS model at finite $\theta$ angle with massive quarks, as in \cite{Bartolini:2016dbk}, to which we adhere for the notation. The action of the model is reviewed in section \ref{revWSS} and it is given by the sum of three terms. The first is the action of $N_f$ D8-branes in Witten's background, eq.~(\ref{SYM}), the second gives the contribution of quark masses (\ref{Smass}) and the third reproduces a non-vanishing $\theta$ term (\ref{SF2}). We focus on the $N_f=1$ case, in which the total low-energy action, once the integration on the four-sphere has been performed, is \cite{Bartolini:2016dbk}
\be\label{Stot1}
S = -\frac{\kappa}{2} \int d^4x \, dz (1+z^2) (\partial_\mu \widehat{A}_z)(\partial^\mu \widehat{A}_z) - \frac{\chi_g}{2} \int d^4x (\theta-\varphi)^2 + 2cm \int d^4x \cos\varphi \,,
\ee
where $\widehat{A}_z$ is the $U(1)$ gauge field living on the D8-brane, depending on the coordinates $x^{\mu}$ of the four-dimensional flat spacetime and on the (dimensionless) holographic coordinate $z$, $m$ is the quark mass, while the relevant scaling of the constants is $\kappa \sim N M_{KK}^2$, $c \sim NM_{KK}^3$ and $\chi_g\sim M_{KK}^4$, and
\be
\varphi \equiv -\frac{1}{\sqrt{2}}\int dz \widehat{A}_z \,.
\ee
In the following we will take $\theta=\pi$. 
Note that with the Ansatz
\be
\label{ans1}
\widehat{A}_z(x^{\mu},z)=\frac{-\sqrt{2}}{\pi (1+z^2)} \left( \pi - \beta(x^{\mu}) \right) \rightarrow \varphi=\pi-\beta(x^{\mu}) \,,
\ee
the action (\ref{Stot1}) reduces to the four-dimensional integral of the chiral Lagrangian for the $\eta'$ meson (identified, up to a constant, with $\beta$, the fluctuation of $\varphi$ around $\pi$)
\be\label{chil1}
{\cal L}=-\frac{\kappa}{\pi} (\partial_\mu\beta) (\partial^\mu\beta) - \frac{\chi_g}{2} \beta^2 - 2cm \cos\beta \,.
\ee
This yields, indeed, the reduction to $\theta=\pi$ and $N_f=1$ of the well-known expression for the $\eta'$ mass
\be
m^2_{\eta'}=m^2_{WV}+m^2_\pi\cos(\theta/N_f) \, .
\ee
The holographic dictionary gives the following expressions for the two contributions to $m_{\eta'}$ \cite{Bartolini:2016dbk}. The first term is the Witten-Veneziano mass \cite{Witten:1979vv,Veneziano:1979ec}, giving the topological contribution
\be
m^2_{WV}=\frac{2N_f \chi_g}{f^2_\pi}=\frac{\pi N_f \chi_g}{2\kappa} \,,
\ee
while the second contribution is proportional to the quark mass\footnote{The case $N_f=1$ is special since pions do not exist. However, we will still refer to $m^2_\pi$ as a physical parameter of the theory. Moreover, since we are working at large $N$ we have that $f^2_{\eta'}=f^2_\pi$.}
\be
m^2_\pi=\frac{4cm}{f^2_\pi}=\frac{\pi cm}{\kappa} \,.
\ee
Since $\theta\in[-\pi,\pi]$, we see that if $N_f>1$ the $\eta'$ is always massive, but if $N_f=1$ and $\theta=\pi$ then the $\eta'$ is massless when $m^2_\pi=m^2_{WV}$. Note that
\be
\frac{m^2_\pi}{m^2_{WV}} \sim N \left( \frac{m}{M_{KK}} \right) \,,
\ee
so in the validity regime of the WSS model we can investigate both the cases $m^2_\pi \gg m^2_{WV}$ (at very large number of colors) and $m^2_\pi \ll m^2_{WV}$ (at very small quark mass), provided they are both much smaller than $M^2_{KK}$. In fact, we should work at $m^2_\pi \ll M^2_{KK}$ in order to neglect higher order corrections to the mass term of the WSS model, and at $m^2_{WV} \ll M^2_{KK}$ in order to neglect the flavor backreaction on Witten's background. Note also that $m^2_{WV}$ scales like $N_f/N$, hence it is indeed suppressed in the large $N$ limit, but should still be taken as a finite leading effect in $N_f$. To make contact with the language of \cite{Gaiotto:2017tne}, $m^2_{WV}$ and $m^2_{\pi}$ have the role of $m_0$ and $m$, respectively, see also figure \ref{GKS}.

We now review the vacuum structure of the theory. The minimum of the potential
\be
V = \frac{\chi_g}{2} \beta^2 + 2cm \cos\beta = \frac{f^2_\pi}{2} \left( \frac{1}{2}m^2_{WV}\beta^2 + m^2_\pi \cos\beta \right) \,,
\ee
is given by $\beta=\pm\beta_0$ (we take $\beta_0>0$) such that
\be
\sin\beta_0=\frac{m^2_{WV}}{m^2_\pi}\beta_0 \,.
\label{eqb0}
\ee
If $m^2_\pi\leq m^2_{WV}$ CP is conserved\footnote{Recall that a CP transformation acts as $\varphi\rightarrow 2\pi-\varphi$, which corresponds to $\beta\rightarrow -\beta$ in the case $\theta=\pi$.} since the only minimum is $\beta_0=0$, and hence there are no domain walls. If $m^2_\pi>m^2_{WV}$ CP is (spontaneously) broken since there are non-vanishing CP-related minima $\beta=\pm\beta_0$, signalling a second order phase transition in the 4d theory at $m^2_{\pi}=m_{WV}^2$. We will now consider two different regimes of the parameters, $M^2_{KK}\gg m^2_\pi \gtrsim m^2_{WV}$ and $M^2_{KK}\gg m^2_\pi \gg m^2_{WV}$.

The first regime corresponds to small $\beta$, where we can approximate the potential up to a quartic term and the minimum is
\be\label{beta1}
\beta_0=\left(6\left(1-\frac{m^2_{WV}}{m^2_\pi}\right)\right)^{1/2} \,.
\ee
In the second regime the minimum is
\be\label{beta2}
\beta_0=\pi\left(1-\frac{m^2_{WV}}{m^2_\pi}\right) \,.
\ee
The domain wall interpolating between the two vacua $-\beta_0$ and $\beta_0$ corresponds to the configuration $\beta(x_1)$ which solves the equation of motion following from (\ref{chil1})
\be\label{eqbeta}
\partial^2 \beta(x_1) = m^2_{WV} \beta(x_1) - m^2_\pi \sin\beta(x_1) ~,
\ee
and interpolates from $-\beta_0$ at $x_1\rightarrow -\infty$ to $+\beta_0$ at $x_1\rightarrow +\infty$. By continuity, the domain wall interpolates between the two vacua passing by $\beta=0$ for $x_1=0$. It is thus clear that while in the first regime the two vacua (\ref{beta1}) are close to each other in field space, since they are close to zero, in the second regime the two vacua are far apart, despite the fact that in the formula (\ref{beta2}) they both seem close to $\pi$ (or $-\pi$, which is the same).

The tension of the domain wall is given by the spatial integral of the on-shell Lagrangian, $T=\int dx_1 {\cal L}$. In the two regimes analyzed above the tension scales as
\be\label{tension1}
T \sim \frac{f^2_\pi}{m^2_\pi}(m^2_\pi-m^2_{WV})^{3/2} \qquad \text {(first regime)} \,,
\ee
and
\be\label{tension2}
T \sim f^2_\pi m_\pi  \qquad \text {(second regime)} \,.
\ee
We note that the transition between the two regimes is smooth. In fact, the former tension (\ref{tension1}) approaches the latter one (\ref{tension2}) in the limit $m^2_\pi\gg m^2_{WV}$, so that there is no 3d phase transition on the domain wall between these two regimes. In particular, increasing the value of $m^2_{\pi}$, always remaining in the region $M^2_{KK} \gg m^2_\pi > m^2_{WV}$, the $\eta'$ trajectory smoothly explores the region from $-\pi$ to $\pi$, interpolating through $\beta=0$. As we are in the case $N_f=1$, the domain-wall theory is the trivial theory of the center of mass d.o.f., which can be entirely described by the $\eta'$  Lagrangian and, remarkably, finds its exact holographic description in the WSS model.

Given the holographic interpretation of the domain wall in the small quark mass regime as a non-trivial profile for the gauge field, we can readily compute its flux as
\be
\frac{1}{\sqrt{2}}\int dx_1 dz {\widehat F}_{x_1 z} = -\int dx_1 \, \partial_{x_1} (\pi-\beta(x_1)) = 2\beta_0  ~,
\label{dissolute}
\ee
realizing the dissolution of the D6-brane in the D8-brane, equation (\ref{D6inD8}), {\it i.e.} the holographic dual of the trivial phase of the domain-wall theory mentioned in \cite{Gaiotto:2017tne} and reviewed above. Note that \eqref{dissolute} can be interpreted as an effective tension for the dissolved D6-brane. Indeed it nicely fits with the fact that the DW tension should go to zero when $m_\pi=m_{WV}$.

Let us now consider the $N_f>1$ case. There the $\eta'$ is never massless and CP is spontaneously broken at $\theta=\pi$ for any value of $m^2_\pi$, and hence the 4d second order phase transition happens at $m^2_{\pi}=0$, in agreement with expectations. Holographically, we have to consider both the non-Abelian $SU(N_f)$ gauge field (dual to the $N^2_f-1$ light pseudo-Goldstone bosons) as well as the Abelian $U(1)$ part $\widehat{A}_z$ (dual to the $\eta'$ meson). In the same way as in the $N_f=1$ case, at finite $\theta$ angle and small quark mass the chiral Lagrangian, which always includes the $\eta'$, is obtained as the low-energy limit of the WSS action. As expected, the small quark mass phase of the domain-wall theory is correctly described by the holographic model as a CP$^{N_f-1}$ non-linear $\sigma$-model. 

Strictly speaking, one should distinguish two regimes. When $m_\pi \ll m_{WV}$, $\eta'$ plays only a spectator role and the DW dynamics is purely determined by the pion chiral Lagrangian. Conversely, when $m_\pi \gg m_{WV}$, the $\eta'$ becomes indistinguishable from the other pions and the DW dynamics should be determined by a $U(N_f)$ chiral Lagrangian. Still, we believe that the same non-linear $\sigma$-model describes the dynamics on the DW. This is actually further motivated by consistency with the phase transition, as we now discuss. As a final remark, note that for $N_f > 2$ the non-trivial configuration (\ref{ans1}) reduces the  CS term $\omega_5(\mathcal{A})$ in the action (\ref{SYM}) to a Wess-Zumino term for the three-dimensional theory, consistently with the expectations in \cite{Komargodski:2017keh,Gaiotto:2017tne}.

\subsection{Large quark mass regime and the phase transition}
\label{phtrs}

Let us now address the large quark mass regime. As we already discussed, we have a rather coherent picture in the small quark mass regime, as well as in the pure YM scenario, in which quarks are decoupled from the theory. In the stringy picture, on one side we have that tachyon condensation causes the dissolution of the D6-branes in the D8-branes, signalled by the presence of a non-trivial flux of the D8 gauge field, which acts as a D6 charge. On the other side, we have a non-trivial $U(1)_N$ TQFT on the D6-branes, which are stable objects located at the tip of the cigar in Witten's background. The tension of the pure YM domain walls is 
\be
T \sim N \, M^3_{KK} \,.
\label{tension3}
\ee
The dependence on the scale $M_{KK}\sim \Lambda$ simply follows from dimensional analysis whereas the scaling with $N$ is a consequence of the YM Lagrangian being proportional to $N$.

Both of these two cases are fully under control in the holographic set-up. Moreover, QCD at very large quark mass ($m\gg\Lambda$) should behave at low energy as pure YM, implying that the domain-wall dynamics should also be the same.  Clearly, the quark mass term as introduced in section {\ref{revWSS} and discussed in the previous section does no longer correctly describe how to implement quark masses in this regime. The D6-branes must in any case be stable objects in this regime, {\it i.e.}~they should not form a bound state with the D8. This in turn entails that the D6/D8 open
string spectrum is not anymore tachyonic, there is no scalar condensation and the D6 world-volume theory can be non-trivial.
 
The different behavior of the domain-wall tension and the different theories living on its world-volume signal a 3d phase transition, which occurs for some finite value $m_{tr}$ of the quark mass, of order $\Lambda$ (or, in the holographic language, when the parameter $m^2_{\pi}$ is order $M^2_{KK}$).

When $m^2_\pi$ approaches $M^2_{KK}$ the low-energy theory descending from the holographic set-up is no more the chiral Lagrangian, since the latter does not correctly capture the relevant degrees of freedom. Even from a pure field-theoretical point of view, when the chiral Lagrangian breaks down the would-be Goldstone bosons are no more the lightest degrees of freedom of the theory. In fact, glueballs (whose masses are of order $\Lambda$ and do not depend on quark mass) become the relevant degrees of freedom.\footnote{Notice however that in the large $N$ holographic limit, the mixing effects between the two sectors are suppressed \cite{Brunner:2016ygk}.}
 At $m\sim\Lambda$ the effective Lagrangian should then include glueballs and other massive mesons, while at larger quark masses we should end up with a theory of glueballs, only.

In the latter regime, the interpolating field whose non-vanishing VEV signals spontaneous CP breaking is no more the $\eta'$ meson (considering the case $N_f=1$), but the lightest pseudoscalar glueball $\tilde{G}$ having the same quantum numbers, associated to the gauge-invariant parity-odd operator ${\rm {Tr}} (F\tilde{F})$.\footnote{More precisely the interpolating field is a linear combination of $\tilde{G}$ and $\eta'$. The coefficients of this linear combination depend on the quark mass, such that for $m\gg\Lambda$ ($\ll \Lambda$), it is approximated by $\tilde{G}$ ($\eta'$). At intermediate value of the quark mass, when we expect the domain-wall phase transition to occur, we have to consider the full linear combination.} Thus, we can also think about $m_{tr}$ as the mass scale where the $\eta'$ mass equals that of $\tilde{G}$, which precisely corresponds to $m^2_{\pi}\sim M^2_{KK}$ in the holographic language. Note that the tension of the domain wall seems to be continuous at the transition. Indeed, using the scaling properties of the parameters of the theory, (\ref{tension2}) behaves like (\ref{tension3}) when $m^2_{\pi}$ approaches $M^2_{KK}$. For greater values of the quark mass the domain-wall tension is ``frozen" to be (\ref{tension3}) instead of growing with $m_{\pi}$ as in (\ref{tension2}), right because quarks can be integrated out and one ends up with a theory of glueballs. This seems to suggest that the phase transition is smooth.

Holographically $m_{tr}$ corresponds to the value of the quark mass such that the D6/D8 open string tachyon becomes massless. A complete computation of the tachyon spectrum in the regime of interest is clearly outside of our approximations. However, we can ask if the (approximatively) constant gauge field near the tip of the cigar in the small quark mass regime gives a positive contribution to the tachyon mass square, thus signalling its tendency to become smaller (in absolute value) as quark mass is increased. Since the non-trivial profile for the D8 gauge field acts as an electric field on the endpoint of a string attached to the brane, it effectively pulls it away from $z=0$. This causes the D6/D8 open string to stretch, suggesting in turn a positive contribution to the mass of the tachyon. Let us show how this works in the small quark mass regime, where our set-up is under full control.

Since the D6-branes are localized at the tip of the cigar, where they intersect the D8-branes, the D6/D8 string is localized at the tip, as well. The metric at the tip of the cigar, see eq.~(\ref{mtip}), is flat and thus, in absence of the gauge field on the D8s, the mass of the tachyon can be deduced from a well-known flat space expression. Adapting it to our set-up we get
\be
m_T^2 = -\frac{\pi}{2} \, T_s \sim - \lambda M_{KK}^2\,,
\ee
where $T_s$ is the confining string tension (\ref{ts}). The gauge field on the D8s, through the string boundary term $\int d\tau \widehat A_z \dot z$, shifts the canonical momentum conjugated to the $z$ coordinate of the string endpoint. The shift affects the mass-shell condition for the string ground state in such a way that - in analogy with what happens in the standard compact case, see {\it e.g.} \cite{Polchinski} - the mass of the tachyon receives a positive contribution of the order
\be
\delta m^2_T \sim M_{KK}^2(\delta{\widehat A}_z)^2_{z=0}\sim \beta_0^2 M_{KK}^2~,
\ee    
where $\delta {\widehat A}_z$ is the variation of the gauge field (\ref{ans1}) in the vacuum around  its value at $m=0$. Thus, the expression above arises by computing the zero-point energy of the string and subtracting the $m=0$ value. The dependence on the mass parameter is hidden in the value of $\beta_0$ as deduced from (\ref{eqb0}). Since $\beta_0$ grows from $0$ to $\pi$ for increasing mass parameters, we see that the desired effect on the tachyon mass shows up: increasing $m$ causes the tachyon squared mass to go towards less negative values. 

The computation above, being strictly valid in the small quark mass regime, can only qualitatively account for a trend and it does not allow us to approach the desired phase transition point, where the tachyon mass is expected to vanish, since $\delta m^2_T/m_T^2\sim 1/\lambda\ll1$. Going to large quark masses in the set-up proposed by \cite{AK, Hashimotomass} would require summing up all open string instanton corrections. We do not know what the result of the resummation will be.  However, the trend we have observed above suggests that, within a complete resummed model, for sufficiently large $m$ the tachyon becomes massless, the D6/D8 bound state is marginal and the DW theory is $U(1)_{N}$ plus a scalar at criticality. Finally, for larger $m$ one can integrate out the tachyon and remain with $U(1)_{N}$ in the IR.

\section {Conclusions and outlook}
\label{conc}

The primary goal of this work was to derive the dynamics of domain walls arising in massive QCD at finite $\theta$ angle by means of the Witten-Sakai-Sugimoto model. Our findings confirm the field theory results derived in \cite{Gaiotto:2017tne}, giving further evidence that the two theories are in the same universality class. Even more interestingly, the string construction gives support to the idea that the domain-wall theory can be fully captured by a three-dimensional matter coupled CS theory, as suggested in \cite{Gaiotto:2017tne}.

Indeed, besides reproducing the expected phases on the domain walls, the string construction provides a concrete realization of the (bosonic version of the) three-dimensional CS theory which was conjectured in \cite{Gaiotto:2017tne} to actually describe the domain-wall dynamics. Note that this theory (its fermionic dual, in fact) falls in the class of CS theories which were proposed in \cite{Komargodski:2017keh} to admit a quantum phase. The WSS model concretely realizes one such example.\footnote{See \cite{Hong:2010sb,Jensen:2017xbs,Armoni:2017jkl} for string theory constructions which addressed directly 3d CS boson-fermion dualities, although in a regime where no quantum phase for the fermionic theory is expected to occur.}

In the WSS model domain walls are described by D6-branes, whose low-energy effective theory is a $U(1)_N$ CS theory with scalar fundamental matter. The tachyonic nature of such scalar fields and their condensation in the low quark mass regime suggest that tachyon dynamics is responsible for the phase transition between the low mass $\sigma$-model phase and the large mass pure $U(1)_N$ CS phase. This gives a natural indication for a phase transition of second order. While we have provided evidence for the positive contribution to tachyon square masses as quark masses are increased, we cannot provide a clear-cut proof that tachyons become massless at the phase transition, since the latter occurs outside the regime of validity of the massive WSS model. Moreover, the effective theory describing tachyon dynamics at the transition might be strongly coupled, making any quantitative analysis difficult.  It would be very interesting to investigate this issue further. 

An indirect argument in favor of the aforementioned tachyon dynamics may come from a $T$-dual set-up (which, as such, is not expected to change the nature of  phase transitions), in which flavor branes could be separated along a Dirichlet direction (along the lines of \cite{Karch:2002sh,Kruczenski:2003uq}) from both domain wall and gauge theory branes. Such a setting would make manifest the vanishing of tachyon masses as quark masses are increased to some critical value.

We have considered mainly single domain walls. As we are in the large $N$ limit, the vacua at $\theta=\pi+2\pi k$ are almost degenerate, and hence we could also consider walls, or interfaces, between vacua further apart. These interfaces would have a similar description in terms of a stack of $k$ D6-branes. The world-volume theory is seemingly also very similar, the first bet being a  $U(k)_N$ CS theory with $N_f$ fundamental scalars. This would lead to the possibility to study large values of $k$, possibly taking into account the backreaction of the D6-branes.

There are several other interesting properties of the 3d theories living on the DWs, as for example the presence of skyrmions and of line operators, for which, as far as we know, a holographic description is not yet known and which would be worth to investigate.

\section*{Acknowledgments} 
 
We are grateful to Adi Armoni, Vladimir Bashmakov, Francesco Benini, Francesco Bonechi, Andrea Cappelli, Paolo Di Vecchia, Zohar Komargodski, Carlo Maccaferri, Flavio Porri and Domenico Seminara for relevant comments and discussions.  A special thank goes to Zohar Komargodski for feedback on a preliminary draft version. 
R.A. and P.N. acknowledge support by IISN-Belgium (convention 4.4503.15) and by the FRS-FNRS under the `Excellence of Science' EOS be.h project n.~30820817, M.B. by the MIUR PRIN Contract 2015 MP2CX4 ``Non-perturbative Aspects Of Gauge Theories And Strings". R.A. is a Research Director of the FRS-FNRS (Belgium). The work of P.N. is financially supported by a scholarship of the International Solvay Institutes. M.B., F.B and A.C. acknowledge partial support by INFN Iniziative Specifiche ST\&FI and GAST. The authors acknowledge the Galileo Galilei Institute for Theoretical Physics for the hospitality during the program ``Supersymmetric Quantum Field Theories in the Non-perturbative Regime".


\appendix

\section{Witten background}
\label{appreview}
In the holographic regime, the background sourced by $N$ D4-branes wrapped on $S_{x_4}$ as described in the main body of the paper is given by \cite{Witten:1998zw}
\bea 
ds^2&=&\left({u\over R}\right)^{3/2} \left(dx^\mu dx_\mu + f(u)d x_4^2 \right)+ \left({R\over u}\right)^{3/2}{du^2\over f(u)}
+R^{3/2}u^{1/2}d\Omega_4^2 \nonumber \\     f(u)&=&1-{u_0^3\over u^3}\ ,  \qquad     e^\Phi=g_s{u^{3/4}\over R^{3/4}}\ , \qquad F_4 = 3 R^3 \omega_4 ~, 
\eea
where $\mu=(0,1,2,3)$, the radial variable $u$ has dimensions of length and ranges in $[u_0,\infty)$, the circle $x_4$ has length $\beta_4=2\pi/M_{KK}$, $R=(\pi g_s N)^{1/3}l_s$ and 
$\omega_4$ is the volume form of $S^4$.  Absence of conical singularities gives the relation $9 u_0 \beta_4^2 = 16 \pi^2 R^3$ and implies that the $(x_4,u)$ space has a cigar topology, with the tip of the cigar at $u=u_0$. 

With the change of variables
\begin{equation}
u^3 = u_{0}^3 + u_{0} {\tilde u}^2\;,\quad \varphi = \frac{2\pi}{\beta_4} x_4 \,,\label{defUcoord}
\end{equation}
and parameterizing the $({\tilde u},\varphi)$ plane in coordinates 
$y=\tilde u\cos\varphi \,,  z=\tilde u \sin \varphi$, where $z\in (-\infty,\infty)$, the cigar metric reads
\begin{equation}
d s^2_{(y,z)} = \frac{4}{9}\left(\frac{R}{u}\right)^{3/2}\left[\left(1-q(\tilde u)z^2\right)d z^2 + \left(1-q(\tilde u)y^2\right)d y^2 - 2zy \, q(\tilde u)d z d y\right]\,,
\end{equation}
where $u$ is given as a function of $z$ and $y$ and $q(\tilde u)$ is defined by $q(\tilde u)= \frac{1}{{\tilde u}^2}\left(1-\frac{u_0}{u}\right)$. In these coordinates the embedding for the antipodal D8-branes is simply given by $y=0$.

It is instructive to see what the metric at the tip of the cigar $u=u_0$ looks like. Using the previous expressions and introducing the dimensionless variables $({\hat z},{\hat y}) = u_0^{-1}(z,y)$ it reads 
\be
ds^2_{\rm{tip}} = 2\pi\alpha' T_s dx_{\mu}dx^{\mu} + \frac94 \frac{2\pi\alpha' T_s}{M_{KK}^2} \left[d{\hat z}^2+d {\hat y}^2 +  d\Omega_4^2\right]\,,
\label{mtip}
\ee
where
\be
\label{ts}
T_s = \frac{2\lambda}{27\pi} M_{KK}^2
\ee
is the confining string tension which can be deduced from the holographic computation of the rectangular Wilson loop. 


\end{document}